\documentclass[twocolumn,epjc3]{svjour3mod}

\usepackage{hyperref}
\usepackage{graphicx,colordvi}
\usepackage{amsmath,amssymb,slashed,cite}
\usepackage{booktabs,tabulary}
\usepackage{dsfont}
\usepackage{color}
\usepackage{epstopdf}

\newcommand{\diff}{\text{d}}
\newcommand{\MeV}{\,\text{MeV}}
\newcommand{\GeV}{\,\text{GeV}}
\renewcommand{\Im}{\text{Im}\,}

\newcommand{\disc}{\text{disc}\,}
\newcommand{\mpi}{M_\pi}
\newcommand{\nl}{\notag\\}

\newcommand{\F}{\mathcal{F}}
\newcommand{\A}{\mathcal{A}}
\newcommand{\efac}{e} 
\newcommand{\beq}{\begin{equation}}
\newcommand{\eeq}{\end{equation}}
\newcommand{\bsp}{\begin{sloppypar}}
\newcommand{\esp}{\end{sloppypar}}

\smartqed  
\RequirePackage{graphicx}

\journalname{Eur. Phys. J. C}
\begin{document}

\title{Dispersive analysis of the pion transition form factor}
\author{M.\ Hoferichter\thanksref{addr1,addr2,addr3}
        \and
        B.\ Kubis\thanksref{addr4}
        \and
        S.\ Leupold\thanksref{addr5} 
        \and
        F.\ Niecknig\thanksref{addr4} 
        \and
        S.\ P.\ Schneider\thanksref{addr4}
}
\institute{Institut f\"ur Kernphysik, Technische Universit\"at
Darmstadt, D--64289 Darmstadt, Germany\label{addr1}
\and
ExtreMe Matter Institute EMMI, GSI Helmholtzzentrum f\"ur
Schwerionenforschung GmbH, D--64291 Darmstadt, Germany\label{addr2}
\and
Albert Einstein Center for Fundamental Physics, Institute for Theoretical Physics,
University of Bern, Sidlerstrasse 5, CH--3012 Bern, Switzerland\label{addr3}
           \and
Helmholtz-Institut f\"ur Strahlen- und Kernphysik (Theorie) and
         Bethe Center for Theoretical Physics,
         Universit\"at Bonn,
         D--53115 Bonn, Germany\label{addr4}
           \and
Institutionen f\"or fysik och astronomi, Uppsala Universitet, Box
516, S--75120 Uppsala, Sweden\label{addr5}
}

\date{}

\maketitle

\begin{abstract}
We analyze the pion transition form factor using dispersion theory.
We calculate the singly-virtual form factor in the time-like region based on data for the $e^+e^-\to 3\pi$ cross section, generalizing previous studies on $\omega,\phi\to3\pi$ decays and $\gamma\pi\to\pi\pi$ scattering,
and verify our result by comparing to $e^+e^-\to\pi^0\gamma$ data.
We perform the analytic continuation to the space-like region, predicting the poorly-constrained space-like transition form factor below $1\GeV$, and extract the slope of the form factor at vanishing momentum transfer $a_\pi=(30.7\pm0.6)\times 10^{-3}$. We derive the dispersive formalism necessary for the extension of these results to the doubly-virtual case, as required for the pion-pole contribution to hadronic light-by-light scattering in the anomalous magnetic moment of the muon.

\end{abstract}

\section{Introduction}
\label{sec:intro}

\bsp
One of the biggest challenges of contemporary particle physics is the unambiguous identification of signs of beyond-the-standard-model 
physics. While high-energy experiments are mainly devoted to the search for new particles, 
high-statistics low-energy experiments can provide such a high precision that standard-model predictions can be seriously 
scrutinized. 
A particularly promising candidate for such an enterprise is the gyro-magnetic ratio of the muon, for a review 
see~\cite{Jegerlehner:2009ry}.
Since the muon is
an elementary spin-1/2 fermion, the decisive quantity is the deviation of its gyro-magnetic ratio $g$ from its classical value.
This difference, caused by quantum effects, is denoted by $(g-2)_\mu$.
\esp

From the theory side the potential to isolate effects of physics beyond the standard model is limited by the 
accuracy of the standard-model prediction. Typically the limiting factor is our incomplete understanding of the 
non-perturbative sector 
of the standard model, i.e.\ the low-energy sector of the strong interaction, which is governed by hadrons as the relevant
degrees of freedom instead of the elementary quarks and gluons. In fact, for $(g-2)_\mu$ the hadronic contributions by far dominate
the uncertainties for the standard-model prediction. The largest hadronic contribution, hadronic vacuum polarization (HVP), enters at order $\alpha^2$ in the fine-structure constant $\alpha=e^2/(4\pi)$ and  can be directly related to 
{\em one} observable quantity, the cross section of the reaction $e^+ e^- \to \,$hadrons, by means of dispersion theory. In that way a reliable error estimate of HVP emerges from the knowledge of the experimental uncertainties in the measured cross section. At order $\alpha^3$ there are next-to-leading-order iterations of HVP as well as a new topology, hadronic light-by-light scattering (HLbL)~\cite{Calmet:1976kd}.
It was recently shown in~\cite{Kurz:2014wya} that even next-to-next-to-leading-order iterations of HVP are not negligible at the level of accuracy required for the next round of $(g-2)_\mu$ experiments planned at FNAL~\cite{Welty-Rieger:2013saa}
 and J-PARC~\cite{Saito:2012zz}, while an estimate of next-to-leading-order HLbL scattering indicated a larger suppression~\cite{Colangelo:2014qya}.

With the increasing accuracy of the cross-section measurement for $e^+ e^- \to \,$hadrons that can be expected in the 
near future~\cite{Blum:2013xva}, 
the largest uncertainty for $(g-2)_\mu$ will then reside in the HLbL contribution. The key quantity here 
is the coupling of two (real or virtual) photons to any hadronic single- or many-body state. This quantity 
is not directly related to a 
single observable. However, it is conceivable to build up the hadronic states starting with the ones most dominant at low energies,
in particular the light one- and two-body intermediate states. 
Based on a dispersive description of the HLbL tensor an initiative has recently been started to relate 
the one- and two-pion contributions for HLbL scattering to observable quantities~\cite{Bern:lbl,roadmap,Bern:lbl-procs}.\footnote{A different approach, based on dispersion relations for the Pauli form factor instead of the HLbL tensor, was recently proposed in~\cite{Pauk:2014rfa}.
For a first calculation in lattice QCD,
an alternative strategy to reduce the model dependence in the HLbL contribution, see~\cite{Blum:2014oka}.}
\bsp
The present work should be understood as an input for this initiative. 
We focus on the lowest hadronic state, the neutral pion,
and its coupling to two (real or virtual) photons (a similar program is currently also being pursued for $\eta$ and $\eta'$, see~\cite{Stollenwerk,Hanhart_eta}). Thus the central object of interest is 
the pion transition form factor. Its importance for the HLbL contribution to $(g-2)_\mu$ 
has been stressed early on, see e.g.~\cite{Bijnens:2007pz,Jegerlehner:2009ry,Czerwinski:2012ry}, and
triggered many studies of the transition from factor in this
context~\cite{Hayakawa:1997rq,Bijnens:2001cq,KN,MV,Nyffeler:2009tw,Goecke,Masjuan,Terschlusen:2013iqa,Roig:2014uja}. 
It is defined by
\begin{align}
  \label{eq:defpiTFF}
  & \int d^4x \, e^{iq_1\cdot x} \, i \langle 0 \vert T \, j_\mu(x) \, j_\nu(0) \vert \pi^0(q_1+q_2) \rangle
  \notag \\
  & = -\epsilon_{\mu\nu\alpha\beta} \, q_1^\alpha \, q_2^\beta \, F_{\pi^0\gamma^*\gamma^*}(q_1^2,q_2^2),
\end{align}
where 
\beq
  \label{eq:defelqcurr}
  j_\mu = e \, \sum\limits_f Q_f \, \bar q_f \gamma_\mu q_f
\eeq
denotes the electromagnetic current carried by the quarks and $Q_f$ the electric charge of the quark of flavor $f$ (in units
of the proton charge $e$). 
\esp

The normalization of the form factor is given by a low-energy theorem~\cite{LET_pi0_1,LET_pi0_2,LET_pi0_3}. In the chiral limit
one finds
\beq
  \label{LET_2pi}
  F_{\pi^0\gamma^*\gamma^*}(0,0) \to \frac{e^2}{4\pi^2 F_\pi} \equiv F_{\pi\gamma\gamma},
\eeq
which agrees with experiment to a remarkable accuracy, see~\cite{Bernstein_review} for a recent review. 
In~\eqref{LET_2pi} $F_\pi = 92.2\MeV$ denotes the pion decay constant~\cite{PDG}. 

\bsp
For the dispersive treatment of the HLbL contribution to $(g-2)_\mu$ as envisaged in~\cite{Bern:lbl,roadmap,Bern:lbl-procs} one needs the pion transition form factor for arbitrary space-like 
virtualities $q_1^2$ and $q_2^2$ of the
two photons. We will approach this aim in a multi-step process. In the present work we will formulate the dispersive framework 
for the general doubly-virtual transition form factor, but restrict the numerical analysis to the singly-virtual case, both in the space- and time-like regions. We will use data on $e^+e^-\to 3\pi$ to fix the parameters and predict the cross section for $e^+e^-\to\pi^0\gamma$ as well as the space-like transition form factor to demonstrate the viability of the approach. While presently low-energy space-like data are scarce~\cite{CELLO,CLEO}, new high-statistics data can be expected in the near future from BESIII (see~\cite{Benayoun:2014tra,Amaryan:2013eja}), which makes a calculation of the space-like singly-virtual form factor particularly timely.
In a second step, the experimental information from $e^+e^-\to\pi^0\gamma$ both in space- and time-like kinematics will then serve as additional input for a full analysis of the doubly-virtual form factor.
\esp

\begin{figure}
\centering
\includegraphics[width=0.35\linewidth,clip]{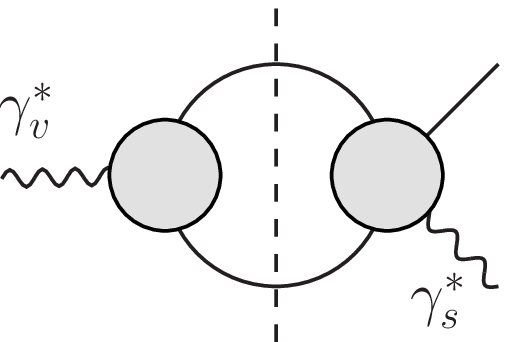}\qquad 
\includegraphics[width=0.35\linewidth,clip]{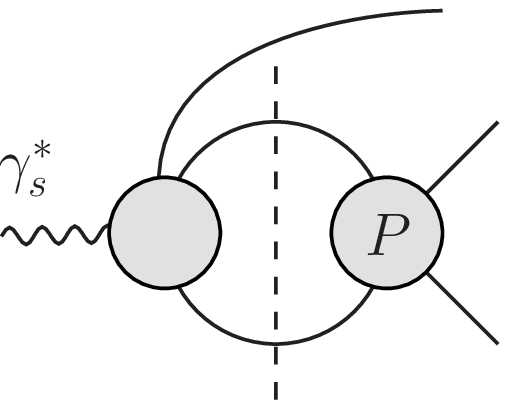} 
\caption{Two-body unitarity relations for $\gamma^*_v\to \gamma^*_s \pi^0$ (left) and the $\gamma^*_s\to3\pi$ amplitude (right). Solid (wiggly) lines denote pions (photons) and the $P$ indicates $P$-wave final-state interactions.} 
\label{fig:TFF_unitarity}
\end{figure}

The basic idea of the dispersive approach for the calculation of the pion transition form factor is 
its reconstruction from the most important intermediate states in the unitarity relation (see also~\cite{Gorchtein,Amaryan:2013eja}).
At low energies these are
the two-pion and three-pion states with isospin $1$ and $0$, respectively. Assuming perfect isospin symmetry one of the two
photons of the $\pi^0\gamma^* \gamma^*$ amplitude must be in an isovector and one in an isoscalar state. We shall denote this
assignment by the indices $v$ and $s$, respectively. Then at low energies the unitarity relation for $\gamma^*_v \gamma^*_s \pi^0$ is
dominated by $\gamma^*_v \to \pi^+ \pi^- \to \gamma^*_s \pi^0$, see the left diagram in Fig.~\ref{fig:TFF_unitarity}. Additional inelasticities start contributing only at an 
invariant mass of the isovector photon above $1\GeV$, predominantly in the form of four pions, cf.~\cite{pionvff_Hanhart}. 
We will not consider such contributions explicitly in the present work, but estimate their potential impact by variations
of the $\pi\pi$ phase shifts in the inelastic region.
The crucial building blocks of the dispersive treatment are the charged pion vector form factor $F_\pi^V$, defined by
\begin{align}
  \langle 0 \vert j^\mu(0) \vert &\pi^+(p_+) \, \pi^-(p_-) \rangle = \nonumber \\ 
  &-e \, (p_+^\mu - p_-^\mu) \; F_\pi^V\!\left((p_++p_-)^2\right),
  \label{eq:defpionvecFF}
\end{align}
and the amplitude for the $\gamma^* \to 3\pi$ reaction. The pion vector form factor with its normalization $F_\pi^V(0) = 1$ 
has been studied in great detail both from the theoretical and experimental
side, see e.g.~\cite{pionvff_BELLE,pionvff_BaBar,pionvff_KLOE,pionvff_Anant,pionvff_Hanhart}. 
It is closely related to the Omn\`es function to which we will come back in Sect.~\ref{sec:3pi},
see also~\cite{omegaTFF,g3pi} for more details.

\begin{figure}
\centering
\raisebox{2mm}{\includegraphics[width=0.35\linewidth,clip]{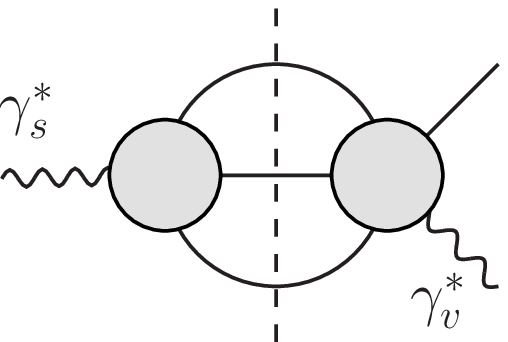}}\qquad 
\includegraphics[width=0.45\linewidth,clip]{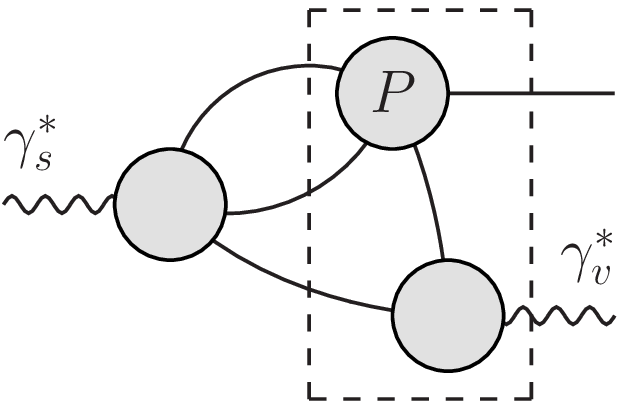} 
\caption{Three-body unitarity relation for $\gamma^*_s\to \gamma^*_v \pi^0$ (left) and the approximation inherent in our formalism (right).} 
\label{fig:TFF_3B_unitarity}
\end{figure}

In contrast, the structure of the amplitude for $\gamma^* \to 3\pi$ is much more involved. It will be discussed in detail
in Sect.~\ref{sec:3pi}. Its two-body unitarity relation, illustrated by the right diagram in Fig.~\ref{fig:TFF_unitarity}, involves 
the rescattering of pion pairs, which can be resummed in terms of the $P$-wave $\pi\pi$ phase shift
within the dispersive approach. While two-body unitarity is exact, we do not consider full three-body unitarity as required by the $3\pi$
intermediate states in $\gamma^*_s \to \pi^+ \pi^-\pi^0 \to \gamma^*_v \pi^0$, see left diagram in Fig.~\ref{fig:TFF_3B_unitarity}. However, with two-body unitarity fully implemented, the $\pi\pi$ rescattering in $\gamma_s^*\to3\pi$ generates topologies such as the one shown in the right diagram in Fig.~\ref{fig:TFF_3B_unitarity}, which manifestly contains three-pion cuts. The part of this diagram indicated by the dashed box can be interpreted as a special case of the full $\pi^+ \pi^-\pi^0 \to \gamma^*_v \pi^0$ amplitude. Therefore, in our framework the structure of the left-hand cut in $3\pi\to\gamma^*\pi$ is approximated by pion pole terms.

The rest of the paper is organized as follows: in Sect.~\ref{sec:3pi} we describe our framework for the determination of the 
$\gamma^*\to3\pi$ amplitude. In Sect.~\ref{sec:disp-general} we formulate the general dispersion relation for 
the pion transition form factor with arbitrary virtualities for the two photons. In Sect.~\ref{sec:pi0g} we specialize 
the general framework to the case of one on-shell and one time-like photon. As a first application we will determine 
the cross section of the reaction $e^+e^-\to\pi^0\gamma$ and compare to the corresponding experimental results. 
Section~\ref{sec:spacelike} is devoted to the analytic continuation into the space-like region as well as the calculation of the slope
of the form factor at zero momentum transfer. The Dalitz decay region is discussed in Sect.\ \ref{sec:Dalitzdec}. 
We close with a summary and outlook in Sect.~\ref{sec:sum}. An Appendix is added to discuss the comparison of our results to 
the simple vector-meson-dominance picture.

\section{The $\gamma^*\to3\pi$ amplitude}
\label{sec:3pi}

\subsection{Formalism}

A key ingredient for the dispersive calculation of the pion transition form factor 
is the amplitude for the reaction $\gamma^*(q) \to \pi^+(p_+) \pi^-(p_-)\pi^0(p_0)$. We define
\begin{align}
  \langle 0 \vert j_\mu(0) \vert &\pi^+(p_+) \, \pi^-(p_-) \, \pi^0(p_0) \rangle = \nonumber \\
  & - \epsilon_{\mu\nu\alpha\beta} \, p_+^\nu \, p_-^\alpha \, p_0^\beta \, \F(s,t,u;q^2)
  \label{eq:defgamma3pion}
\end{align}
with $q=p_++p_-+p_0$, $s=(p_++p_-)^2$, $t=(p_-+p_0)^2$, $u=(p_++p_0)^2$, and $s+t+u=3\mpi^2+q^2$.

The low-energy limit of $\F$ is dictated by the chiral
anomaly. In the chiral limit this leads to the identification~\cite{WZW_1,WZW_2,WZW_3,WZW_4,WZW_5,g3pi}
\beq
  \label{eq:defF3pi}
  \F(0,0,0;0) \to \frac{e}{4\pi^2 F_\pi^3}\equiv F_{3\pi}.
\eeq
A comment is in order to which extent the chiral predictions~\eqref{LET_2pi} and~\eqref{eq:defF3pi} have been confronted with experiment so far. 
$F_{\pi\gamma\gamma}$ has been tested up to $1.5\%$ in Primakoff measurements of $\pi^0\to\gamma\gamma$~\cite{PrimEx} including chiral~\cite{Bijnens:1988kx,Goity:2002nn} and radiative~\cite{Ananthanarayan:2002kj} corrections, the former up to two-loop order~\cite{Kampf:2009tk}.
Both the world average~\cite{PDG} and the PrimEx result~\cite{PrimEx} are fully consistent with the chiral tree-level prediction~\eqref{LET_2pi}, the former even at $1\%$ accuracy, while chiral corrections predict an increase of up to $2\%$ mainly due to $\pi^0\eta$ mixing~\cite{Kampf:2009tk}, in slight tension with the world average. Here, we use~\eqref{LET_2pi} directly, given that apart from the very low-energy region the associated uncertainties are sub-dominant.     

In contrast to this high accuracy the  
extractions of $F_{3\pi}$ both from Primakoff measurements~\cite{Antipov} (with chiral and radiative corrections from~\cite{Bijnens90,Hannah,Ametller}) and $\pi^- e^-\to \pi^- e^- \pi^0$~\cite{Scherer} presently allow a test at the $10\%$ level only. In~\cite{g3pi} a dispersive framework (see also~\cite{Hannah,Holstein,Truong} for earlier work in this direction) was presented  that provides a two-parameter description of the $\pi^-\gamma\to\pi^-\pi^0$ cross section valid up to $1\GeV$. This opens the possibility to profit from the high-statistics Primakoff data currently analyzed at COMPASS~\cite{Friedrich:2014dna} concerning the extraction of $F_{3\pi}$ to higher accuracy.

\bsp
We decompose $\F$ as 
\beq
\label{eq:ee3piAmp}
\F(s,t,u;q^2) = \F(s,q^2) + \F(t,q^2) + \F(u,q^2). 
\eeq
This decomposition neglects discontinuities in $F$- and higher partial waves, see~\cite{Hannah}.
Using the ($s$-channel) partial-wave decomposition 
\begin{align}
\label{eq:pwdec}
\F(s,t,u;q^2) &=\sum_{\ell\;{\rm odd}}f_\ell(s,q^2)P'_\ell(\cos\theta_s),\notag\\
\cos\theta_s &= \frac{t-u}{\kappa(s,q^2)},\notag\\
\kappa(s,q^2) &= \sigma_\pi(s)\lambda^{1/2}(q^2,\mpi^2,s),
\end{align}
with the K\"all\'en function $\lambda(x,y,z)=x^2+y^2+z^2-2(xy+yz+xz)$ and $\sigma_\pi(s)=\sqrt{1-4\mpi^2/s}$,
we find that the function $\F(s,q^2)$ in~\eqref{eq:ee3piAmp} is related to the $P$-wave amplitude according to~\cite{V3pi}
\begin{align}
f_1(s,q^2) &= \F(s,q^2) + \hat \F(s,q^2),\notag\\
\hat\F(s,q^2) &=  \frac{3}{2}\int_{-1}^1\diff z\big(1-z^2\big) \F\big(t(s,q^2,z),q^2\big),
\label{eq:detf1}
\end{align}
with 
\begin{align}
  \label{eq:deftfs}
  t(s,q^2,z) = \frac12 \, (3\mpi^2+q^2-s) + \frac12 \, \kappa(s,q^2) \, z  \,.
\end{align}
Note that for positive $q^2$ the evaluation of~\eqref{eq:pwdec} is straightforward, while some care is needed for the proper
analytic continuation of the square roots for negative $q^2$. Therefore the framework presented here can be immediately applied
for instance to the singly-virtual time-like transition form factor, as will be shown in Sect.~\ref{sec:pi0g}. For the
corresponding space-like form factor, to be tackled in Sect.~\ref{sec:spacelike}, we will refrain from an analytic 
continuation of the formulae presented here but instead use a dispersion relation to determine the space-like transition form factor from the 
imaginary part of the time-like one. 
\esp

For fixed $q^2$, the quantity $\F(s,q^2)$, given in~\eqref{eq:detf1}, only has a right-hand cut starting at $s=4\mpi^2$. 
The left-hand cut of the 
partial wave $f_1(s,q^2)$ entirely resides in $\hat\F(s,q^2)$.  
Furthermore, the amplitude develops a three-pion cut for $q^2 > 9\mpi^2$, i.e.\ in kinematics allowing for 
the physical decay $\gamma^* \to 3\pi$.  In this situation, the right- and left-hand cuts in $s$ begin to overlap,
which leads to a significant complication of the analytic structure, see the corresponding discussion in~\cite{V3pi}.

\bsp
The discontinuity of the partial wave $f_1(s,q^2)$ along the right-hand cut is given by 
\beq
\label{eq:Watson}
 \disc f_1(s,q^2) = 2i\,f_1(s,q^2)\theta(s-4\mpi^2)\sin\delta(s)e^{-i\delta(s)},
\eeq
where $\delta(s)\equiv\delta_1^1(s)$ is the $\pi\pi$ $P$-wave phase shift. 
Noting that $\disc f_1(s,q^2)=\disc\F(s,q^2)$ along the right-hand cut, we can recast this relation
into the form
\begin{align}\label{eq:unrel}
 \disc\F(s,q^2) &= 2i\,\bigl(\F(s,q^2)+\hat\F(s,q^2)\bigr)\notag\\
& \qquad \times \theta(s-4\mpi^2)\sin\delta(s)e^{-i\delta(s)}.
\end{align}
A once-subtracted dispersive representation solving~\eqref{eq:unrel} is given by~\cite{V3pi}
\begin{align}
\label{eq:FOmnes}
 \F(s,q^2)&=\Omega(s) \\
 & \times \bigg\{a(q^2) + \frac{s}{\pi}\int_{4\mpi^2}^\infty \diff s'\frac{\hat \F(s',q^2)\sin\delta(s')}{s'(s'-s)|\Omega(s')|}\bigg\},\notag
\end{align}
where 
\beq
\Omega(s) = \exp\biggl\{\frac{s}{\pi}\int_{4\mpi^2}^\infty \diff s'\frac{\delta(s')}{s'(s'-s)}\biggr\}
\eeq 
is the Omn\`es function~\cite{Omnes}.
\esp

An important property of~\eqref{eq:FOmnes} concerns its linearity in the subtraction function $a(q^2)$, which follows from the fact that
$\hat \F$ is defined in terms of the angular average of $\F$ itself~\eqref{eq:detf1}. In this way, $a(q^2)$ 
takes the role of a normalization, so that in practice~\eqref{eq:detf1} and~\eqref{eq:FOmnes} are solved by iteration for $a(q^2)\to1$, while the full solution is recovered by multiplying with $a(q^2)$ in the end. However, since $t$ as a function of $s$ implicitly depends on $q^2$, the subtraction function is not the only source of $q^2$ dependence in the full solution.

For fixed virtualities $q^2 = M_\omega^2,\,M_\phi^2$ the solutions of~\eqref{eq:detf1} and \eqref{eq:FOmnes} have been studied in~\cite{V3pi} to describe the vector-meson decays $\omega,\,\phi\to3\pi$.\footnote{For a variant of this calculation see~\cite{Danilkin:2014cra}.} In this case the respective 
subtraction constant 
$a$ is fixed by the overall normalization of the Dalitz plot distribution
and hence the corresponding partial decay width. 
The main complication when extending~\eqref{eq:FOmnes}
to arbitrary virtualities $q^2$ of the incoming photon arises from the fact that $a$ depends on
$q^2$, a dependence that cannot be predicted within the dispersive framework itself, but has to be determined by 
different methods. 
Physically, $a(q^2)$ contains the information how the isoscalar photon couples to hadrons. At low energies, this coupling is dominated by
the three-pion state and can be accessed in $e^+e^-\to 3\pi$. For the extraction of $a(q^2)$ we need a representation that
preserves analyticity and accounts for the phenomenological finding that the three-pion state 
is strongly correlated to the very narrow $\omega$ and $\phi$ resonances.
We take
\beq
\label{eq:BW}
a(q^2)=\alpha + \beta q^2 +\frac{q^4}{\pi}\int_{s_\text{thr}}^\infty \diff s'\frac{\Im \A(s')}{{s'}^2(s'-q^2)},
\eeq
with $\Im \A$ modeled using two relativistic Breit--Wigner functions 
\begin{align}
\label{eq:A}
\A(q^2)&=\frac{c_\omega}{M_\omega^2-q^2-i \sqrt{q^2} \Gamma_\omega(q^2)} \nl
& +\frac{c_\phi}{M_\phi^2-q^2-i\sqrt{q^2} \Gamma_\phi(q^2)}.
\end{align}
In the following we refer to $\Im \A$ as the spectral function. 
In~\eqref{eq:A} $\Gamma_{\omega/\phi}(q^2)$ is the energy-dependent width of the $\omega/\phi$ meson, respectively. 
We take into account the main decay channels of $\omega$ and $\phi$ via
\begin{align}
\Gamma_\omega(q^2) &= \frac{\gamma_{\omega\to3\pi}(q^2)}{\gamma_{\omega\to3\pi}(M_\omega^2)}\Gamma_{\omega\to3\pi}
+ \frac{\gamma_{\omega\to\pi^0\gamma}(q^2)}{\gamma_{\omega\to\pi^0\gamma}(M_\omega^2)}\Gamma_{\omega\to\pi^0\gamma}, \nl
\Gamma_\phi(q^2) &= \frac{\gamma_{\phi\to3\pi}(q^2)}{\gamma_{\phi\to3\pi}(M_\phi^2)}\Gamma_{\phi\to3\pi} \nl
&+
\sum \limits_{K=K^+,K^0}
\frac{\gamma_{\phi\to K\bar K}(q^2)}{\gamma_{\phi\to K\bar K}(M_\phi^2)}\Gamma_{\phi\to K\bar K}, 
\label{eq:allthewidths}
\end{align}
where $\Gamma_i$ denotes the measured partial decay width for the decay $i$, while the energy-dependent coefficients are
given by 
\begin{align}
\gamma_{\omega\to\pi^0\gamma}(q^2) &= \frac{(q^2-\mpi^2)^3}{(q^2)^{3/2}}, \nl
\gamma_{\phi\to K\bar K}(q^2) &= \frac{(q^2-4M_K^2)^{3/2}}{q^2},
\end{align}
and the calculation of $\gamma_{\omega/\phi\to3\pi}(q^2)$ is performed along the lines described in~\cite{V3pi}. 
For completeness we also include the $\pi^0\gamma$ decay channel of the $\omega$, which strictly speaking corresponds to a radiative correction.
As a consequence the threshold $s_\text{thr}$ in~\eqref{eq:BW} is actually $M_{\pi^0}^2$ instead of $9\mpi^2$.
However, we checked that as expected the impact of the $\pi^0\gamma$ channel is very small numerically.

The representation~\eqref{eq:BW} can be understood as a dispersively improved Breit--Wigner 
parametrization~\cite{Lomon,Moussallam:gg*pipi}: the reconstruction of the real part via a dispersive
integral ensures a reasonable behavior of the phase of $a(q^2)$ despite the energy dependence of the widths.
We decide to subtract~\eqref{eq:BW} twice:
the first subtraction constant $\alpha$ is fixed by the chiral anomaly for $\gamma\to3\pi$ at the real-photon point (corrected for quark-mass renormalization)~\cite{Bijnens90,g3pi},
\beq
  \label{eq:alphaF3pi}
  \alpha=\frac{F_{3\pi}}{3}\times (1.066 \pm 0.010) \equiv \alpha_{3\pi}.
\eeq
The second subtraction $\beta$ serves as an additional background term and is fitted to 
$e^+e^-\to3\pi$ cross-section data, together with the residues $c_\omega$ and $c_\phi$.
Note that the precise form of the spectral function in~\eqref{eq:A} is irrelevant: 
the only requirement is to have an analytically rigorous representation of the cross section. 

Finally, we give the explicit relation between the $\gamma^*(q) \to \pi^+(p_+) \pi^-(p_-)\pi^0(p_0)$ amplitude~\eqref{eq:ee3piAmp}
and the $e^+e^-\to3\pi$ cross section (neglecting the electron mass)
\beq
  \label{eq:epemcross1}
  \sigma_{e^+ e^- \to 3\pi} = \int_{s_\text{min}}^{s_\text{max}} \diff s \int_{t_\text{min}}^{t_\text{max}} \diff t \,
  \frac{\diff^2\sigma}{\diff s \, \diff t}, 
\eeq
with
\beq
  \label{eq:epemcrossdiff}
  \frac{\diff^2\sigma}{\diff s \, \diff t} = 
  \frac{e^2 \, P}{96 \, (2\pi)^3 \, q^6} \, \vert \F(s,t,u;q^2) \vert^2
\eeq
and
\begin{align}
\label{eq:defP}
  P &\equiv  - g^{\mu \mu'} \, \epsilon_{\mu\nu\alpha\beta} \, p_+^\nu \, p_-^\alpha \, p_0^\beta \, 
  \epsilon_{\mu'\nu'\alpha'\beta'} \, p_+^{\nu'} \, p_-^{\alpha'} \, p_0^{\beta'}  \notag \\
  & =  \frac{1}{4} \, (stu-M_\pi^2 \, (q^2-M_\pi^2)^2) \notag \\
  & =  \frac{1}{16} \, s \, \kappa(s,q^2)^2 \, \sin^2\theta_s,
\end{align}
as well as integration boundaries
\beq
  \label{eq:bound1}
  s_\text{min} = 4 M_\pi^2, \qquad s_\text{max} = \big(\sqrt{q^2}-M_\pi \big)^2,
\eeq
and
\begin{align}
 \label{eq:bound2}
 t_\text{min/max} &= (E_-^*+E_0^*)^2\notag\\
 &-\bigg( \sqrt{E_-^{*2}-M_\pi^2} \pm  \sqrt{E_0^{*2}-M_\pi^2} \bigg)^2, \notag\\
 E_-^*&=\frac{\sqrt{s}}{2},\qquad E_0^*=\frac{q^2-s-M_\pi^2}{2\sqrt{s}}.
\end{align}
We note in passing that for fixed, but arbitrary $q^2$ we can predict the shape of the two-fold differential distribution~\eqref{eq:epemcrossdiff}. The knowledge of $a(q^2)$ is only needed for the overall normalization, not for the $s$ and $t$
dependence. 

It has been noted in~\cite{V3pi} that the amplitude representation~\eqref{eq:FOmnes} is not accurate
enough to give a statistically valid description of the very precise $\phi\to 3\pi$ Dalitz plot determination
by the KLOE collaboration~\cite{KLOE:phi}.  For this purpose, a second subtraction was introduced, leading
to the representation 
\begin{align}
 \F(s,q^2)&= \Omega(s) \, \bigg\{a(q^2) + b(q^2)\,s \nl
&+ \frac{s^2}{\pi}\int_{4\mpi^2}^\infty \diff s'\frac{\hat \F(s',q^2)\sin\delta(s')}{{s'}^2(s'-s)|\Omega(s')|}\bigg\} \,
\label{eq:FOmnes2}
\end{align}
(only used for $q^2=M_\phi^2$ in~\cite{V3pi}).
Similarly, for $\gamma\pi\to\pi\pi$ a twice-subtracted amplitude representation was envisaged 
theoretically in~\cite{g3pi}.
For general $q^2$, the second subtraction $b$ will again be $q^2$-dependent.
Provided future measurements allow us to determine such a second subtraction both from 
$\gamma\pi\to\pi\pi$ cross-section data ($b(0)$) and from an $\omega\to3\pi$ Dalitz plot
($b(M_\omega^2)$), the three data points---together with $b(M_\phi^2)$---should permit 
a smooth interpolation of $b(q^2)$ in a representation similar to~\eqref{eq:BW} 
(with only a single subtraction).
In the absence of such additional high-precision data, we will utilize the singly-subtracted 
representation~\eqref{eq:FOmnes} of the $\gamma^*\to 3\pi$ partial wave for the purpose of this study.

\subsection{Fits to $e^+e^-\to3\pi$}

Before turning to the fit results, we first summarize the various uncertainty estimates that we have performed in the context of our fits to $e^+e^-\to 3\pi$. First of all, in the calculation of $\F(s,q^2)$ we used three different $\pi\pi$ phase shifts, the phases from~\cite{CCL,Madrid} and a version of~\cite{CCL} that includes the $\rho'(1450)$ and the $\rho''(1700)$ resonances in an elastic approximation to try to mimic the possible impact of $4\pi$ inelasticities~\cite{V3pi}. In addition, we varied the cutoff $\Lambda_{3\pi}$ in the dispersive integral~\eqref{eq:FOmnes} above which asymptotic behavior is assumed between $1.8$ and $2.5\GeV$, see~\cite{omegaTFF}.

Next, our representation for $a(q^2)$ is only adequate below $1.1\GeV$, given that above this energy excited states of $\omega$ and $\phi$ may contribute. The isoscalar vector resonances listed in~\cite{PDG} below $1.8\GeV$ with a sizable $3\pi$ branching fraction are the $\omega'(1420)$ and the $\omega''(1650)$, with masses and widths
\begin{align}
M_{\omega'}&=(1.425\pm0.025)\GeV,\notag\\
\Gamma_{\omega'}&=(0.215\pm0.035)\GeV,\notag\\
M_{\omega''}&=(1.67\pm0.03)\GeV,\notag\\
\Gamma_{\omega''}&=(0.315\pm0.035)\GeV.
\end{align}
To estimate the effect of these states, we also consider a version of the fits where additional terms for $\omega'$ and $\omega''$ are included in~\eqref{eq:A}, identical to the expression for the $\omega$ apart from the $\pi^0\gamma$ channel (we assume $100\%$ branching fraction to $3\pi$ for $\omega'$ and $\omega''$). In total, we thus have a three- (five-)parameter representation to be fit to data, with free parameters $\beta$, $c_\omega$, $c_\phi$ (and $c_{\omega'}$, $c_{\omega''}$).

The prime source of $e^+e^-\to 3\pi$ data below/above $1.4\GeV$ are the SND~\cite{ee3pi_SND_1,ee3pi_SND_2} and CMD2~\cite{ee3pi_CMD2_1,ee3pi_CMD2_2}/the BaBar data sets~\cite{ee3pi_BaBar}, respectively. 
Restricting the fit (without $\omega'$ and $\omega''$) to the energy region below $1.1\GeV$, we observed that the SND data set can be described with a reduced $\chi^2$ close to $1$, while the CMD2 scans can only be accommodated with a significantly worse $\chi^2$ (around $2.4$). We also checked if the respective fit reproduced the correct chiral anomaly by including $\alpha$ in~\eqref{eq:BW} as another fit parameter. For SND we indeed obtain $\alpha=(1.5\pm 0.2)\alpha_{3\pi}$, while the fit to CMD2 even produces a negative value of $\alpha$.

One explanation for this apparent tension could be provided by the fact that radiative corrections were not treated in exactly the same way in both experiments. Moreover, the CMD2 scans were restricted to a relatively narrow region around the $\omega$ and $\phi$ masses, limiting the sensitivity to the low-energy region (and thus particularly to the chiral anomaly). Such inconsistencies in the $3\pi$ data base were already observed in~\cite{Hagiwara:2011af} in the context of the HVP contribution to $(g-2)_\mu$, where the $3\pi$ channel entered with a global reduced $\chi^2$ of $3.0$. For the present study we will therefore consider two data sets: first, SND+BaBar and, second, the compilation from~\cite{Hagiwara:2011af}, in the following denoted by HLMNT. It includes all data sets mentioned so far as well as some older
experiments~\cite{Cordier:1979qg,Dolinsky:1991vq,Antonelli:1992jx,Akhmetshin:1995vz,Akhmetshin:1998se,ee3pi_CMD2_3}. 
The rationale for doing so is that for the reasons explained above SND/BaBar appear to be the most comprehensive single data sets for low/high energies. Confronting the outcome of fits to the combination of both and to the comprehensive data compilation of~\cite{Hagiwara:2011af} should allow for a reasonable estimate of the impact of the uncertainties in the $e^+e^-\to 3\pi$ cross section on the prediction for the pion transition form factor.

\begin{figure*}
\includegraphics[width=0.498\linewidth,clip]{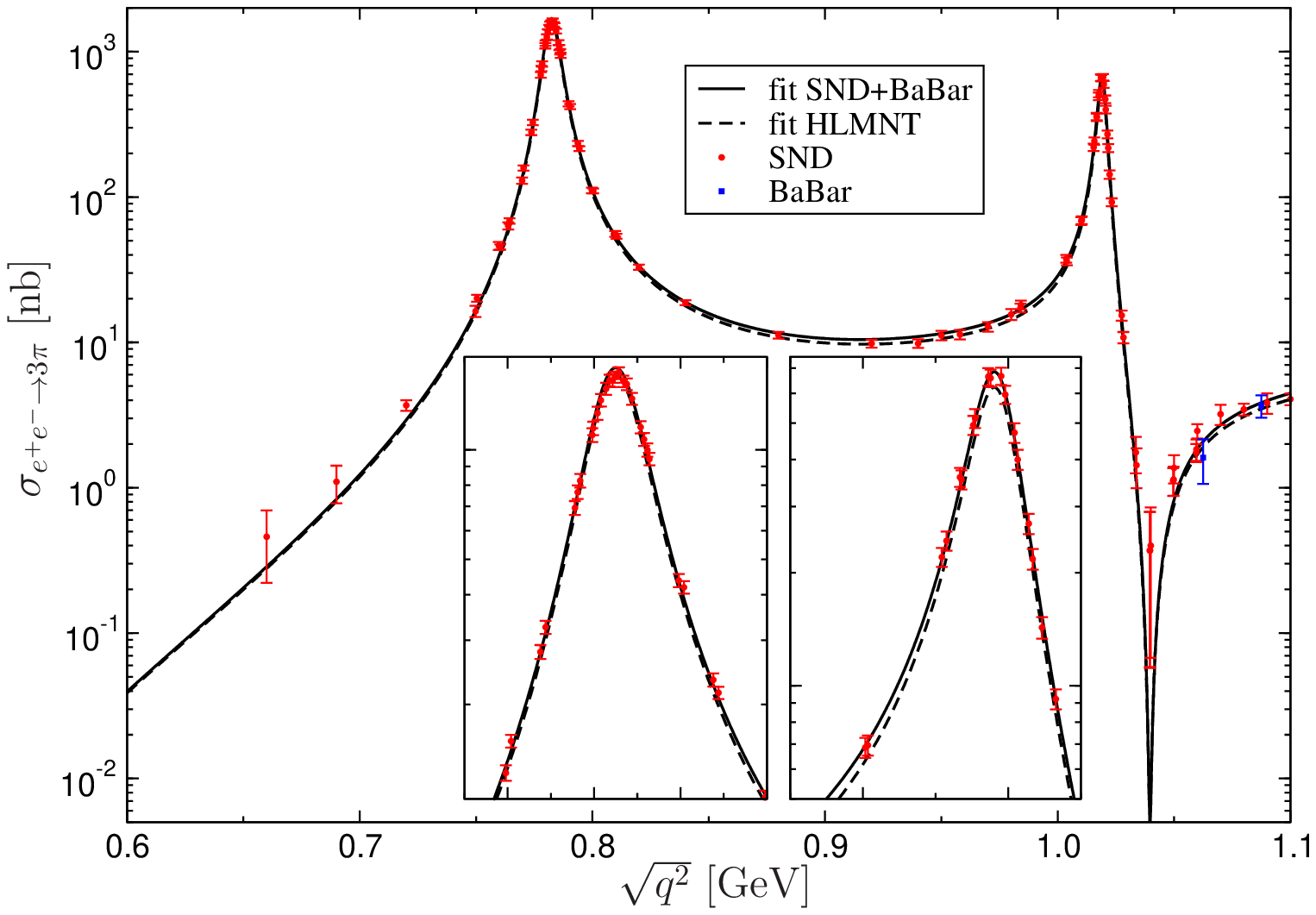} 
\includegraphics[width=0.498\linewidth,clip]{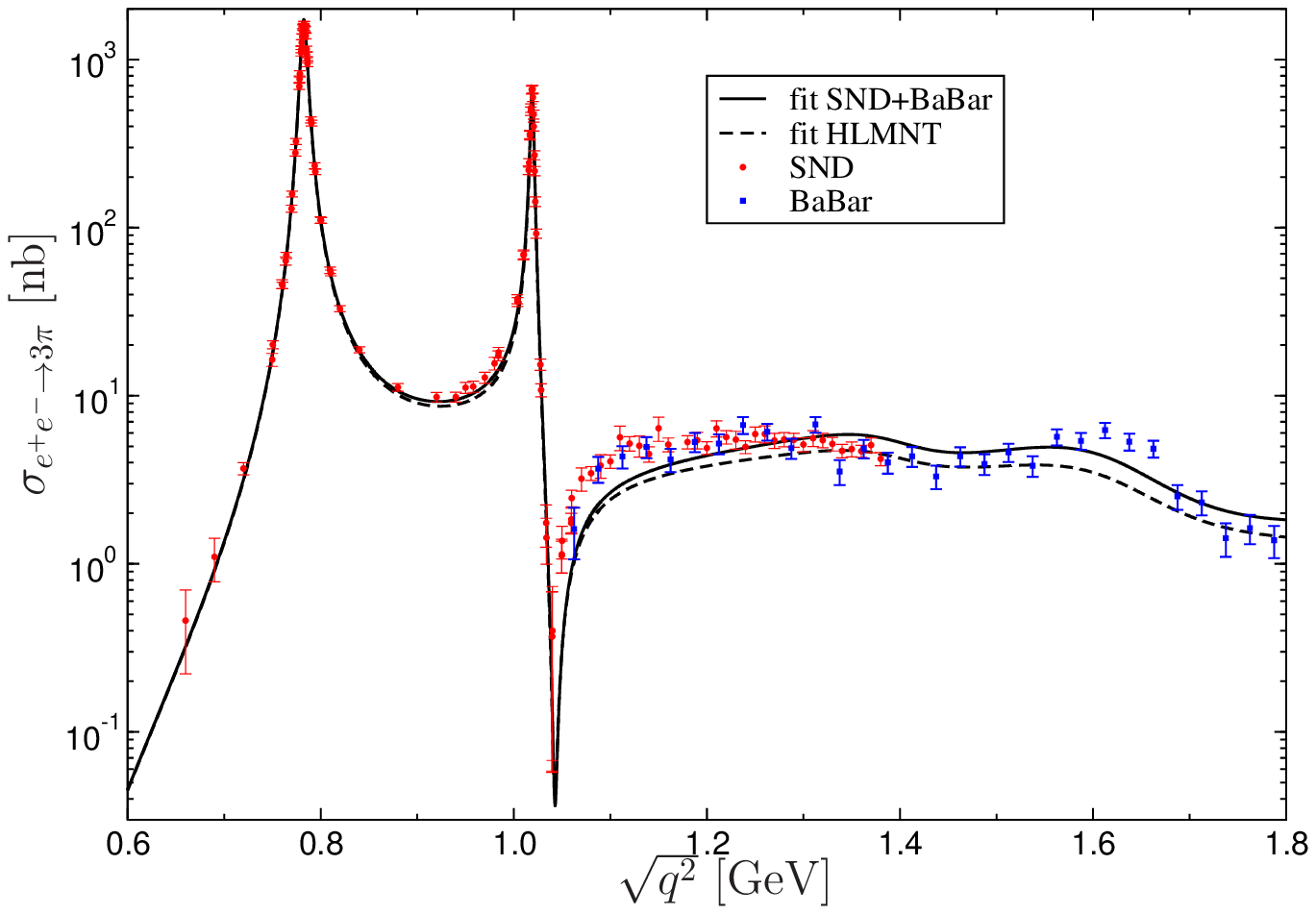} 
\caption{Fit to the $e^+e^-\to3\pi$ cross-section data of~\cite{ee3pi_SND_1,ee3pi_SND_2} and~\cite{ee3pi_BaBar} below $1.1\GeV$ (left) and $1.8\GeV$ (right), with $\pi\pi$ phase shift from~\cite{CCL} and $\Lambda_{3\pi}=2.5\GeV$. The small inserts amplify the regions around the $\omega$ and $\phi$ resonance peaks. Only the fit in the right panel includes $\omega',\omega''$ in the spectral function. The dashed line indicates the outcome of the fit to the data base of~\cite{Hagiwara:2011af}.} 
\label{fig:3pi}
\end{figure*}

\begin{table*}
\centering
\renewcommand{\arraystretch}{1.3}
\begin{tabular}{lcccccc}
\toprule
& $\beta \ [\text{GeV}^{-5}]$ & $c_\omega \ [\text{GeV}^{-1}]$ & $c_\phi \ [\text{GeV}^{-1}]$ & $c_{\omega'} \ [\text{GeV}^{-1}]$ & $c_{\omega''} \ [\text{GeV}^{-1}]$ & $\chi^2/\text{dof}$\\
\midrule
SND+BaBar, $1.1\GeV$ & $5.94\ldots 6.21$ & $2.88\ldots2.90$ & $-(0.392\ldots0.406)$ & --- & --- & $1.01\ldots 1.04$\\
HLMNT, $1.1\GeV$ & $5.92\ldots6.18$ & $2.81\ldots2.83$ & $-(0.374\ldots0.387)$ & --- & --- & $6.33\ldots 6.36$\\
SND+BaBar, $1.8\GeV$ & $7.73\ldots7.78$ & $2.92\ldots2.95$ & $-(0.386\ldots0.400)$ & $-(0.27\ldots0.43)$ & $-(0.70\ldots1.22)$ & $3.18\ldots 3.48$\\
HLMNT, $1.8\GeV$ & $7.78\ldots7.82$ & $2.88\ldots2.90$ & $-(0.366\ldots0.378)$ & $-(0.19\ldots0.32)$ & $-(0.53\ldots1.02)$ & $7.28\ldots 7.62$\\
\bottomrule
\end{tabular}
\renewcommand{\arraystretch}{1.0}
\caption{Fit parameters and reduced $\chi^2$ for the $e^+e^-\to3\pi$ fits to SND+BaBar~\cite{ee3pi_SND_1,ee3pi_SND_2,ee3pi_BaBar} and HLMNT~\cite{Hagiwara:2011af} as described in the main text. The ranges indicate the variation found for the different $\pi\pi$ phase shifts and values of $\Lambda_{3\pi}$.}
\label{tab:3pifit}
\end{table*}
\bsp
The result of the three-parameter fit to SND+BaBar below $1.1\GeV$ is shown in the left panel of Fig.~\ref{fig:3pi}, with fit parameters summarized in Table~\ref{tab:3pifit}. Since the fits to $e^+e^-\to 3\pi$ are hardly distinguishable visually, we only show the curves for the phase shift from~\cite{CCL} and $\Lambda_{3\pi}=2.5\GeV$, but give the ranges for the fit parameters found in the full calculation. For these data sets and energy region the reduced $\chi^2$ is very close to $1$. As alluded to above, the $\chi^2$ deteriorates substantially when fitting to the full data base of~\cite{Hagiwara:2011af}, but the central values of the fit parameters remain largely unaffected.    
\esp

Extending the fit to higher energies by including $\omega'$ and $\omega''$ in the spectral function yields a reasonable fit up to $1.8\GeV$, at the expense of a slight deterioration of the data description between the $\phi$ and $1.2\GeV$, see the right panel of Fig.~\ref{fig:3pi} and Table~\ref{tab:3pifit}. Again, we observe that the fit result is relatively insensitive to the data set chosen, with larger differences evolving in the $\omega',\omega''$ region.
We will use the outcome of this extended fit to estimate the impact of the high-energy region on the analytic continuation of the transition form factor into the space-like region in Sect.~\ref{sec:spacelike}.

\section{Dispersion relations for the doubly-virtual $\pi^0$ transition form factor}
\label{sec:disp-general}

\bsp
We decompose the pion transition form factor into definite isospin components according to
\beq
\label{eq:isospin}
F_{\pi^0\gamma^*\gamma^*}(q_1^2,q_2^2)=F_{vs}(q_1^2,q_2^2)+(q_1\leftrightarrow q_2),
\eeq
where the first/second index refers to isovector ($v$) and isoscalar ($s$) quantum numbers of the photon with momentum $q_1$/$q_2$.
For fixed isoscalar virtuality we can write a once-subtracted dispersion relation in the isovector virtuality~\cite{g3pi}
\begin{align}
\label{eq:DR1}
F_{vs}(s_1,s_2)&=F_{vs}(0,s_2)\\
&+\frac{\efac \, s_1}{12\pi^2}\int^\infty_{4\mpi^2}\diff s'\frac{q_{\pi}^3(s')F_\pi^{V*}(s')f_1(s',s_2)}{s'^{3/2}(s'-s_1)},\notag
\end{align}
where $q_\pi(s)=\sqrt{s/4-M_\pi^2}$, and $F_\pi^{V}(s)$ is the pion vector form factor~\eqref{eq:defpionvecFF}.
Assuming both $F_\pi^{V}(s)$ and $f_1(s,s_2)$ to asymptotically fall off like $1/s$~\cite{Lepage:1980fj,Duncan:1979hi,Froissart:1961ux,V3pi,omegaTFF} 
(for fixed $s_2$), there is a sum rule for the subtraction function in~\eqref{eq:DR1},
\beq
F_{vs}(0,s_2) = 
\frac{\efac}{12\pi^2}\int^\infty_{4\mpi^2}\diff s'\frac{q_{\pi}^3(s')}{s'^{3/2}}F_\pi^{V*}(s')f_1(s',s_2) \,.
\label{eq:SR}
\eeq
This sum rule formally converges only with a partial wave $f_1(s,q^2)$ based on the 
singly-subtracted representation~\eqref{eq:FOmnes}, with a second subtraction~\eqref{eq:FOmnes2}
it can at best be evaluated below a certain cutoff.
The representation~\eqref{eq:DR1} as well as the sum rule~\eqref{eq:SR} have been employed before:
for $s_2=M_{\omega/\phi}^2$, they yield the vector meson transition form factors for $\omega/\phi\to\pi^0\gamma^*$,
including (from the sum rule) the normalization for the real-photon decays~\cite{omegaTFF}.
For $s_2=0$, one obtains the isovector part of the singly-virtual $\pi^0$ transition form factor, with 
the sum rule yielding $F_{\pi\gamma\gamma}/2$~\cite{g3pi}. Numerically, these sum rules were found to be saturated at the $90\%$ level~\cite{omegaTFF,g3pi}.
\esp

Taken together, \eqref{eq:DR1} and \eqref{eq:SR} are equivalent to an unsubtracted dispersion relation
\begin{align}
  F_{vs}(s_1,s_2)=
  \frac{\efac}{12\pi^2}\int^\infty_{4\mpi^2}\diff s'\frac{q_{\pi}^3(s')F_\pi^{V*}(s')f_1(s',s_2)}{s'^{1/2}(s'-s_1)}.
\label{eq:DRunsub}
\end{align}

We can perform a (necessarily less explicit) subtraction of~\eqref{eq:DR1} in $s_2$ as well, 
defining a subtracted partial wave
\beq
\bar f_1(s,q^2) = \frac{f_1(s,q^2)-f_1(s,0)}{q^2}.
\eeq
The alternative formulation of the dispersive representation, making use of the sum rule~\eqref{eq:SR}, then reads
\begin{align}
\label{eq:DR2}
F_{vs}(s_1,s_2)&=F_{vs}(s_1,0) + F_{vs}(0,s_2) - \frac{F_{\pi\gamma\gamma}}{2} \\
&+\frac{\efac \, s_1 \, s_2}{12\pi^2}\int^\infty_{4\mpi^2}\diff s'\frac{q_{\pi}^3(s')F_\pi^{V*}(s')\bar f_1(s',s_2)}{s'^{3/2}(s'-s_1)}.\notag
\end{align}

\section{Time-like form factor and $e^+e^-\to\pi^0\gamma$}
\label{sec:pi0g}

We now specialize the general expressions~\eqref{eq:isospin} and~\eqref{eq:DRunsub} to the singly-virtual case
for further phenomenological investigation. 
The $\pi^0\to\gamma^*\gamma$ transition form factor can be written out explicitly according to
\begin{align}
\label{pig*g}
 F_{\pi^0\gamma^*\gamma}(q^2,0) &=F_{\pi\gamma\gamma}
+\frac{\efac}{12\pi^2}\int_{4M_\pi^2}^\infty \diff s'\frac{q_\pi^3(s')F_\pi^{V*}(s') }{s'^{3/2}} \nl
&\hspace{-35pt} \times \bigg\{f_1\big(s',q^2\big)-f_1(s',0) 
 +\frac{q^2}{s'-q^2}f_1(s',0)\bigg\}.
\end{align}
Here we have again made use of the sum rule~\eqref{eq:SR} to fix the full transition form factor at $q^2=0$ to the chiral anomaly $F_{\pi\gamma\gamma}$. 
Neglecting the mass of the electron for simplicity, 
the relation between the cross section $\sigma_{e^+e^- \to \pi^0 \gamma}$ and the pion transition form factor is given by
\beq
\label{eq:cross-sec-pig}
\sigma_{e^+e^- \to \pi^0 \gamma} = \frac{e^2 \, (q^2-M_{\pi^0}^2)^3}{96\pi \, q^6} \, \vert F_{\pi^0\gamma^*\gamma}(q^2,0) \vert^2.
\eeq

\begin{figure*}
\includegraphics[width=0.498\linewidth, clip]{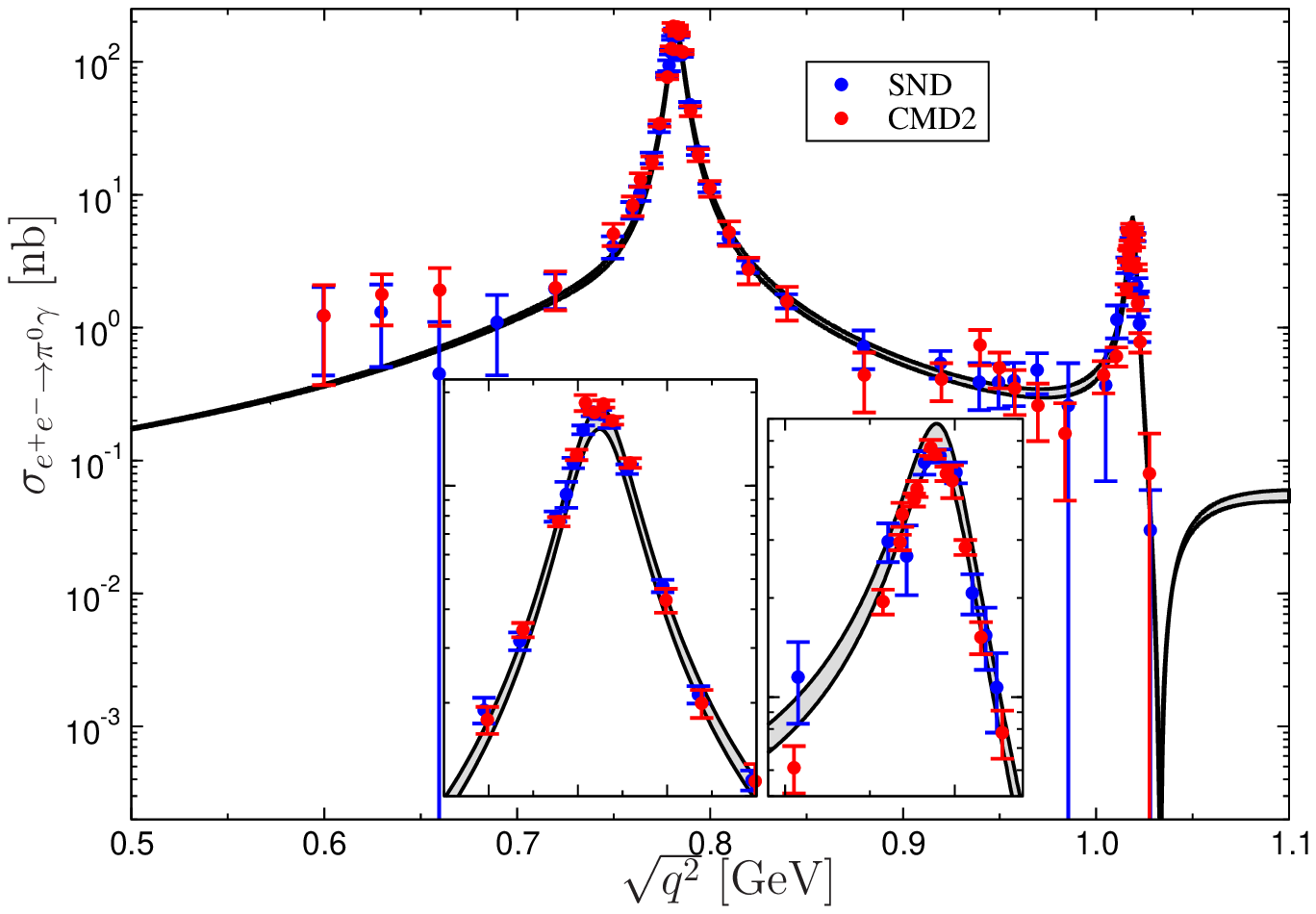}
\includegraphics[width=0.498\linewidth, clip]{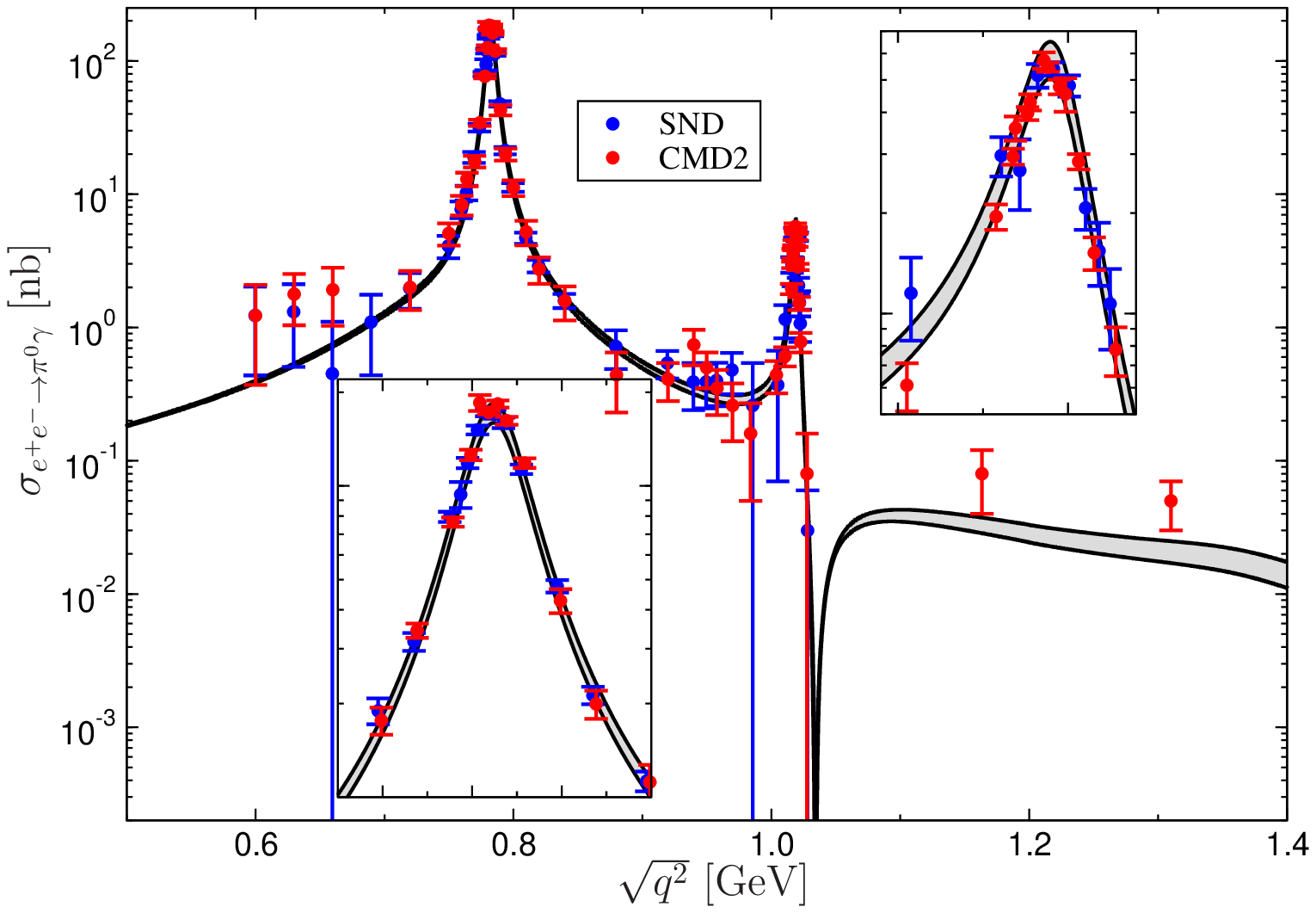}
\caption{$e^+e^-\to\pi^0\gamma$ cross section predicted from $e^+e^-\to3\pi$ (left: fit with $\omega$ and $\phi$ only, right: fit including $\omega'$, $\omega''$), compared to the data of~\cite{eepi0g_SND_1,eepi0g_SND_2,eepi0g_CMD2}. The inserts again zoom in on the $\omega$ and $\phi$ resonance peaks.
The error band represents the variation found by changing the $\pi\pi$ phase shifts, $\Lambda_{3\pi}$, and the $e^+e^-\to 3\pi$ data base as described in the main text.} 
\label{fig:pi0g}
\end{figure*}

\bsp
To ensure consistency with the calculation of the $\gamma^*\to3\pi$ amplitude we assume asymptotic behavior of $F_\pi^V$ and $f_1$ in~\eqref{pig*g} above $\Lambda_{3\pi}$
and use a twice-subtracted Omn\`es representation for $F_\pi^{V}$ (cf.~\cite{Guo:2008nc})
\beq
F_\pi^V(s)=\exp\Bigg\{\frac{\langle r^2\rangle_\pi^V}{6}s+\frac{s^2}{\pi}\int_{4M_\pi^2}^{\infty}\diff s'\frac{\delta(s')}{s'^2(s'-s)}\Bigg\},
\eeq
with a radius $\langle r^2\rangle_\pi^V\sim 0.435\,\text{fm}^2$ and the same phase shift as in the respective version of $f_1(s,q^2)$. 
The isoscalar part, corresponding to the difference $f_1(s',q^2)-f_1(s',0)$ in~\eqref{pig*g}, 
is then calculated by the same methods as in~\cite{V3pi} with the normalization fixed from $e^+e^-\to 3\pi$ as described in Sect.~\ref{sec:3pi}. The isovector part, corresponding to the last term in~\eqref{pig*g}, is completely determined by $f_1(s,0)$ and can thus be measured in $\gamma\pi\to\pi\pi$. Here, we use a finite matching point of $1.2\GeV$ and fix the normalization to the chiral anomaly~\cite{g3pi}, but this representation can be improved once the COMPASS data for $\gamma\pi\to\pi\pi$ become available.  
\esp

\begin{table}
\centering
\renewcommand{\arraystretch}{1.3}
\begin{tabular}{lccc}
\toprule
& SND & CMD2 & SND+CMD2\\
\midrule
$\chi^2/\text{dof}$ & $1.74$ & $4.50$ & $3.12$\\
& $1.05$ & $2.37$ & $1.71$\\\midrule
$\tilde \chi^2/\text{dof}$ & $0.71$ & $1.42$ & $1.06$\\
& $0.56$ & $1.02$ & $0.79$\\
\bottomrule
\end{tabular}
\renewcommand{\arraystretch}{1.0}
\caption{Reduced $\chi^2$ and $\tilde \chi^2$ for the comparison of our result to the $e^+e^-\to\pi^0\gamma$ data of SND~\cite{eepi0g_SND_1,eepi0g_SND_2} and CMD2~\cite{eepi0g_CMD2} as well as the combined data set. In each case, the upper line refers to the fit with $\omega$ and $\phi$ only, the lower line to the fit including $\omega'$, $\omega''$. $\chi^2$ and $\tilde \chi^2$ are calculated for all data points below $1.1\GeV$ (upper line) and $1.4\GeV$ (lower line), respectively.}
\label{tab:epempi0g}
\end{table}

Our result for the $e^+e^-\to\pi^0\gamma$ cross section is shown in Fig.~\ref{fig:pi0g}. 
We repeat the calculation for each set of $\pi\pi$ phase shifts and $\Lambda_{3\pi}$, fitting the isoscalar part in each case both to SND+BaBar and HLMNT.  
The error band in Fig.~\ref{fig:pi0g} represents the uncertainty deduced from scanning over the input quantities in this way. 
Within uncertainties, the outcome agrees perfectly with the $e^+e^-\to\pi^0\gamma$ cross section measured by~\cite{eepi0g_SND_1,eepi0g_SND_2,eepi0g_CMD2}.
We would like to stress that this result is a prediction solely based on the input quantities described above, most prominently, $e^+e^-\to3\pi$ cross-section data, the $\pi\pi$ $P$-wave phase shift, the pion vector form factor, and the low-energy theorems for $F_{3\pi}$ and $F_{\pi\gamma\gamma}$. 

To provide a quantitative measure of the agreement between our result and experiment, we first give the reduced $\chi^2$ of the mean of our band when comparing to the various data sets, see Table~\ref{tab:epempi0g}. However, the usual $\chi^2$ does not account for the theory uncertainty, so that it is not surprising that values significantly larger than $1$ are obtained. If one assumed the theory band to be statistically distributed with mean values $y_\text{th}(q_i)$ and uncertainties $\sigma_\text{th}(q_i)$, uncorrelated for each data point $q_i=\sqrt{q_i^2}$, one could consider the difference between theory and experiment $y_\text{th}(q_i)-y_i$ with combined error $\sqrt{\sigma_\text{th}^2(q_i)+\sigma_i^2}$ and test the distribution for consistency with zero, leading to a modified $\chi^2$,
\beq
\label{modchisqr}
\chi^2\to\tilde\chi^2=\sum_{i=1}^N\frac{\big(y_i-y_\text{th}(q_i)\big)^2}{\sigma_i^2+\sigma_\text{th}^2(q_i)}.
\eeq
The corresponding values for this quantity are also summarized in Table~\ref{tab:epempi0g}. Given that in practice correlations between different points of the theory band are not negligible, the statistical interpretation of~\eqref{modchisqr} is not obvious. However, taken together with the observation that curves within the theory band can be constructed with even smaller $\chi^2$, it provides quantitative evidence for the consistency of our result with the $e^+e^-\to\pi^0\gamma$ data. In addition, the comparison of the $\chi^2$ and $\tilde \chi^2$ for the two fits reveals that, while the $e^+e^-\to3\pi$ fit is deteriorated mostly in the energy region above the $\phi$, including $\omega'$, $\omega''$ improves the agreement with $e^+e^-\to\pi^0\gamma$ below $1.1\GeV$.

\section{Slope parameter and space-like form factor}
\label{sec:spacelike}

We reconstruct the $\pi^0$ transition form factor in the space-like region
again dispersively, making use of the imaginary part determined from the study
of the time-like region in the previous sections
\beq 
\label{disp_TFF}
F_{\pi^0\gamma^*\gamma}(q^2,0) = F_{\pi\gamma\gamma} + \frac{q^2}{\pi}\int_{s_\text{thr}}^\infty \diff s' 
\frac{\Im F_{\pi^0\gamma^*\gamma}(s',0)}{s'(s'-q^2)}.
\eeq
If we assume the transition form factor to fulfill even an unsubtracted dispersion relation,
this relation implies a sum rule for the chiral anomaly
\beq
\label{SR_Fpigg}
F_{\pi\gamma\gamma}=\frac{1}{\pi}\int_{s_\text{thr}}^\infty \diff s' 
\frac{\Im F_{\pi^0\gamma^*\gamma}(s',0)}{s'}.
\eeq
The slope of the form factor obeys
\begin{align}
\label{SR_slope}
a_\pi&=\frac{M_{\pi^0}^2}{F_{\pi\gamma\gamma}}\frac{\partial}{\partial q^2}F_{\pi^0\gamma^*\gamma}(q^2,0)\bigg|_{q^2=0}\notag\\
&=\frac{M_{\pi^0}^2}{F_{\pi\gamma\gamma}}\frac{1}{\pi}\int_{s_\text{thr}}^\infty \diff s' 
\frac{\Im F_{\pi^0\gamma^*\gamma}(s',0)}{s'^2}.
\end{align}

For the evaluation of these relations we need to specify how to treat the high-energy region of the integrals.  Perturbative QCD in the factorization framework of~\cite{Lepage:1980fj} predicts an asymptotic behavior 
\beq
F_{\pi^0\gamma^*\gamma}(-Q^2,0)\sim \frac{2\, e^2 \, F_\pi }{Q^2}.
\eeq
Since the imaginary part has to vanish at least as fast as the real part, we will assume $\Im F_{\pi^0\gamma^*\gamma}(s,0)\sim 1/s$ above a cutoff $\Lambda_{\pi^0}$ and estimate the sensitivity to the asymptotic region by varying $\Lambda_{\pi^0}=(1.1\ldots 1.8)\GeV$. We also considered a constant imaginary part above $\Lambda_{\pi^0}$, finding only moderate shifts, but given that such a behavior contradicts~\cite{Lepage:1980fj} we will not include the corresponding variation in the uncertainty bands shown below. 
Finally, we checked that~\eqref{disp_TFF} indeed reproduces the real part in the time-like region, which is non-trivial in view of the imaginary parts generated by three-body cuts in the calculation of the $\gamma^*\to 3\pi$ amplitude.  

\begin{table}
\centering
\renewcommand{\arraystretch}{1.3}
\begin{tabular}{lcc}
\toprule
& SND+BaBar & HLMNT\\
\midrule
fit below $1.1\GeV$ &  $30.4\ldots 31.2$ & $30.1\ldots 30.9$\\
$\Lambda_{\pi^0}=1.1\GeV$ & $0.989\ldots 1.021$ & $0.976\ldots 1.008$\\\midrule
fit below $1.8\GeV$ &  $30.6\ldots 31.4$ & $30.4\ldots 31.2$\\
$\Lambda_{\pi^0}=1.1\GeV$ & $0.992\ldots 1.026$ & $0.985\ldots 1.019$\\\midrule
fit below $1.8\GeV$ &  $30.4\ldots 31.2$ & $30.3\ldots 31.1$\\
$\Lambda_{\pi^0}=1.4\GeV$ & $0.959\ldots 0.987$ & $0.962\ldots 0.990$\\\midrule
fit below $1.8\GeV$ &  $30.3\ldots 31.1$ & $30.2\ldots 31.0$\\
$\Lambda_{\pi^0}=1.8\GeV$ & $0.944\ldots 0.966$ & $0.947\ldots 0.970$\\
\bottomrule
\end{tabular}
\renewcommand{\arraystretch}{1.0}
\caption{Slope parameter and chiral anomaly from the sum rules~\eqref{SR_Fpigg} and~\eqref{SR_slope}. For each fit and data set the upper line refers to the slope in units of $10^{-3}$, while the lower line gives the sum-rule value for $F_{\pi\gamma\gamma}$ normalized to~\eqref{LET_2pi}. The ranges correspond to the uncertainty due to the $\pi\pi$ phase shift and $\Lambda_{3\pi}$.}
\label{tab:slope}
\end{table}

We first turn to the sum rules for $a_\pi$ and $F_{\pi\gamma\gamma}$, with results summarized in Table~\ref{tab:slope}. For $a(q^2)$ determined from the $e^+e^-\to3\pi$ fit below $1.1\GeV$, including only $\omega$ and $\phi$ in the spectral function, we find the results given in the first two lines for the slope and the chiral anomaly, respectively. For this fit it does not make sense to increase $\Lambda_{\pi^0}$ beyond $1.1\GeV$, given that the fit range in $e^+e^-\to3\pi$ was restricted to this energy region. To estimate the sensitivity to the high-energy region of the dispersive integral, the rest of the table shows the results for the extended fit including in addition $\omega'$ and $\omega''$, with three different values for $\Lambda_{\pi^0}$. For each set of parameters we give the ranges corresponding to the variation of the $\pi\pi$ phase shift and $\Lambda_{3\pi}$ as described in  Sect.~\ref{sec:3pi}.
We find very stable results even for the chiral anomaly, whose sum rule is fulfilled at $5\%$ accuracy, although being more sensitive to high energies (it would not converge if we assumed a constant behavior for the imaginary part above $\Lambda_{\pi^0}$). Averaging over the various fits and data sets we obtain for the slope parameter 
\beq
\label{slope}
a_\pi=(30.7\pm0.6)\times 10^{-3},
\eeq
where the error includes the uncertainties from the $\pi\pi$ phase shift, the cutoffs $\Lambda_{3\pi}$ and $\Lambda_{\pi^0}$, the $e^+e^-\to3\pi$ data sets, and the high-energy contribution to the sum rule (estimated via the $\omega',\omega''$ fits). 
Our result is appreciably more precise than the value $a_\pi=(32\pm4)\times 10^{-3}$ quoted in~\cite{PDG}, which is dominated by a monopole fit to the CELLO data~\cite{CELLO}, or an extraction from an even wider range of space-like data using Pad\'e approximants, $a_\pi=(32.4\pm1.2_\text{stat}\pm1.9_\text{sys})\times 10^{-3}$~\cite{Masjuan}.

Along the same lines, we can also determine the next term in the expansion around $q^2=0$,
\begin{align}
\label{slope2}
b_\pi&=\frac{M_{\pi^0}^4}{F_{\pi\gamma\gamma}}\frac{1}{2}\frac{\partial^2}{\partial (q^2)^2}F_{\pi^0\gamma^*\gamma}(q^2,0)\bigg|_{q^2=0}\notag\\
&=\frac{M_{\pi^0}^4}{F_{\pi\gamma\gamma}}\frac{1}{\pi}\int_{s_\text{thr}}^\infty \diff s' 
\frac{\Im F_{\pi^0\gamma^*\gamma}(s',0)}{s'^3}\notag\\
&=(1.10\pm 0.02)\times 10^{-3},
\end{align}
again with a smaller uncertainty than e.g.\ $b_\pi = (1.06\pm0.09_\text{stat}\pm0.25_\text{sys})\times 10^{-3}$
from~\cite{Masjuan}.
For a comparison of these numbers to the prediction of vector meson dominance~\cite{sakurai}, see~\ref{app:VMD}.

\begin{figure}
\includegraphics[width=\linewidth, clip]{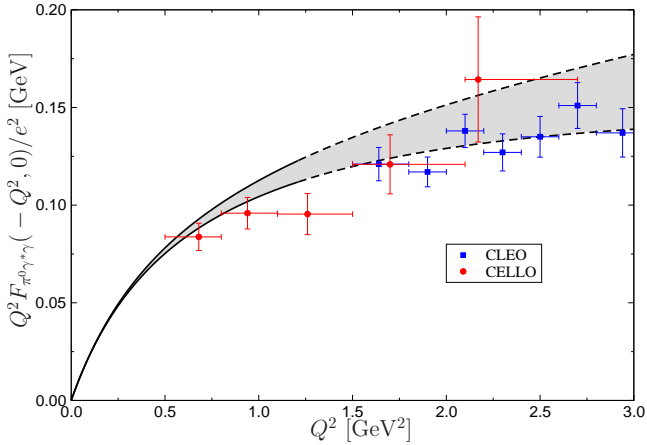}
\caption{Singly-virtual pion transition form factor in the space-like region, compared to CELLO~\cite{CELLO} and CLEO~\cite{CLEO} data.} 
\label{fig:TFF_spacelike}
\end{figure}

Finally, we use~\eqref{disp_TFF} to perform the analytic continuation into the space-like region, see Fig.~\ref{fig:TFF_spacelike}. We follow the convention of the experimental publications to plot $Q^2F_{\pi^0\gamma^*\gamma}(-Q^2,0)/e^2$. In the case of the CELLO data~\cite{CELLO}, provided in the original paper for the form factor without the additional factor of $Q^2$, we use the averages $\langle Q^2\rangle$ given for each bin in the conversion. We also follow the convention to depict the error of the form factor only, and not to propagate an additional uncertainty from the bin size. 

As expected, our prediction for the space-like form factor is very accurate at low energies (better than $5\%$ for $Q^2\leq (1.1\GeV)^2$), while the uncertainties become more sizable above $1\GeV$,
reflecting the limited energy range used as input for the time-like calculation.
The corresponding error band shown in Fig.~\ref{fig:TFF_spacelike} comprises the same uncertainty estimates already discussed in the context of the slope parameter (the energy region $Q^2\geq (1.1\GeV)^2$, which is not reliably described any more in the time-like region, is indicated by the dashed lines in Fig.~\ref{fig:TFF_spacelike}). At low energies the error band is dominated by the variation in the $\pi\pi$ phase shift and $\Lambda_{3\pi}$,\footnote{At very low energies corrections to the low-energy theorem~\eqref{LET_2pi} will become relevant, since the transition form factor is normalized to $F_{\pi\gamma\gamma}$. The corresponding uncertainties are not included in Fig.~\ref{fig:TFF_spacelike}, but due to~\eqref{disp_TFF} can simply be recovered by adding a term $Q^2\Delta F_{\pi\gamma\gamma}/e^2$.} whereas above $1\GeV$ the treatment of the high-energy region in the dispersive integral becomes increasingly important.
The resulting curve is consistent with the existing data base, and will soon be tested by the forthcoming high-statistics low-energy data from BESIII.

\section{Dalitz decay region $\pi^0 \to e^+ e^- \gamma$}
\label{sec:Dalitzdec}

\bsp
So far we have not discussed the third kinematically accessible region of the singly-virtual transition form factor besides $q^2 > M_{\pi^0}^2$ and $q^2 < 0$, i.e.\ the region of the Dalitz decay $\pi^0 \to e^+ e^- \gamma$ with 
$4 m_e^2 < q^2 < M_{\pi^0}^2$, where $m_e$ denotes the electron mass. It is common practice to normalize the corresponding 
partial decay width to the two-photon decay. The normalized differential decay width is given by~\cite{Landsberg}
\begin{align}
  \label{eq:dalitzlandsb}
  \frac{\diff \Gamma_{\pi^0 \rightarrow e^+ e^- \gamma}}{\diff q^2 \, \Gamma_{\pi^0 \rightarrow 2\gamma}} 
  &= \frac{e^2}{6 \pi^2} \, \frac{1}{q^2} \, \sqrt{1 - \frac{4 m_e^2}{q^2}} \left(1 + \frac{2m_e^2}{q^2} \right) \nonumber \\
  &\times
  \left(1 - \frac{q^2}{M_{\pi^0}^2} \right)^3 \bigg| \frac{F_{\pi^0 \gamma^* \gamma}(q^2,0)}{F_{\pi \gamma \gamma}} \bigg|^2.
\end{align}
Absent high-quality data for this differential decay width we just present our result for the integrated
one. In this region of very low momenta it is sufficient to use a polynomial approximation 
for the transition form factor,
\beq
\frac{F_{\pi^0 \gamma^* \gamma}(q^2,0)}{F_{\pi \gamma \gamma}} \approx 1 + a_\pi \frac{q^2}{M_{\pi^0}^2} + b_\pi \frac{q^4}{M_{\pi^0}^4}.
\eeq
Using~\eqref{slope} and~\eqref{slope2} the result is 
\beq
  \label{eq:dalitzdecaywidthours}
   \frac{\Gamma_{\pi^0 \rightarrow e^+ e^-\gamma}}{\Gamma_{\pi^0 \rightarrow 2\gamma}}  = 
   (1.18754 \pm 0.00005) \cdot 10^{-2},  
\eeq
in excellent agreement with the experimental value~\cite{PDG}
\beq
  \label{eq:dalitzdecaywidthexp}
   \frac{\Gamma_{\pi^0 \rightarrow e^+ e^-\gamma}}{\Gamma_{\pi^0 \rightarrow 2\gamma}} \bigg\vert_{\rm exp} = 
   (1.188 \pm 0.035) \cdot 10^{-2}.
\eeq
Value and uncertainty in~\eqref{eq:dalitzdecaywidthours} only reflect our form factor calculation and disregard the issue of radiative corrections~\cite{Kampf:2005tz}.
The impact of the quadratic $b_\pi$ term is  $+2$ in the last digit in~\eqref{eq:dalitzdecaywidthours}.
Note that a pure QED calculation without any form factor yields
\beq
  \label{eq:dalitzdecaywidthQED}
   \frac{\Gamma_{\pi^0 \rightarrow e^+ e^-\gamma}}{\Gamma_{\pi^0 \rightarrow 2\gamma}} \bigg\vert_{\rm no \; FF} = 
   1.18514 \cdot 10^{-2},
\eeq
so that the impact of the transition form factor on the integrated decay width is on the level of $0.2\%$. High-precision data for 
the differential decay width~\eqref{eq:dalitzlandsb} will soon become available in the context of dark-photon searches in $\pi^0\to A'\gamma$ at NA48/2~\cite{Amaryan:2013eja}, but due to the limited sensitivity to the form factor will not improve the PDG value for the slope.
\esp

\section{Summary and outlook}
\label{sec:sum}

We presented the dispersive formalism to analyze the general doubly-virtual pion transition form factor.
This includes all effects from elastic $\pi\pi$ rescattering exactly through the respective phase shifts.  
To determine the isoscalar part that is dominated by $3\pi$ intermediate states, we used data on $e^+e^-\to 3\pi$. 
Furthermore, chiral low-energy theorems on the anomalies $F_{3\pi}$ and $F_{\pi\gamma\gamma}$ were implemented.
As a first step, we carried out the phenomenological analysis of the singly-virtual case. We performed a detailed error analysis and verified our calculation in the time-like region by comparing to data for $e^+e^-\to\pi^0\gamma$, yielding very good agreement between theory and experiment. As further applications of the framework, we provided a precise value for the slope parameter, $a_\pi=(30.7\pm0.6)\times10^{-3}$, as well as for the curvature term, $b_\pi=(1.10\pm0.02)\times10^{-3}$.
Finally, analytic continuation allowed for a prediction for the transition form factor in the low-energy space-like region that should be compared to the upcoming precise BESIII data.

\bsp
To extend the calculation to higher energies requires additional input.  One could for instance match to the predictions of quark counting rules~\cite{Lepage:1980fj}, Regge theory~\cite{Gorchtein}, or light-cone sum rules~\cite{Khodjamirian:1997tk,Agaev:2010aq}. In the time-like region, with consistency between $e^+e^-\to3\pi$ and $e^+e^-\to\pi^0\gamma$ demonstrated, one could also fit simultaneously to both reactions to potentially decrease the uncertainties. 
The most important future extension will concern the generalization to the doubly-virtual case.
This can be applied to predict the leptonic neutral pion decay $\pi^0\to e^+e^-$, but most importantly,
will help pin down the pion-pole contribution to hadronic light-by-light scattering in $(g-2)_\mu$. 
Work in this direction is in progress.
\esp

\begin{acknowledgements}
\bsp
We would like to thank Simon Eidelman, Denis Epifanov, Evgueni Goudzovski, Tord Johansson, Bastian Knippschild, Andrzej Kup\'s\'c, Christoph Redmer, and Thomas Teubner for helpful discussions and correspondence, and Thomas Teubner for making the 
$e^+e^-\to3\pi$ data compilation from~\cite{Hagiwara:2011af} available to us.
Financial support by
BMBF ARCHES, the Helmholtz Alliance HA216/EMMI, the Swiss National Science Foundation,
the DFG (SFB/TR 16, ``Subnuclear Structure of Matter''),
by DFG and the NSFC through funds provided to the Sino-German CRC 110 
``Symmetries and the Emergence of Structure in QCD,''
and by the project ``Study of Strongly Interacting Matter'' 
(HadronPhysics3, Grant Agreement No.~283286) 
under the 7th Framework Program of the EU
is gratefully acknowledged.
\esp
\end{acknowledgements}

\appendix

\section{Narrow-width approximation and comparison to vector meson dominance (VMD)}
\label{app:VMD}

Within the narrow-width approximation we may replace~\cite{g3pi}
\begin{align}
f_1(s',0)F_\pi^{V*}(s') &\to \frac{1}{2}F_{3\pi}\pi\frac{M_\rho^3}{\Gamma_\rho}\delta(s'-M_\rho^2) \,, \nl
f_1(s',q^2)F_\pi^{V*}(s') &\to \frac{3}{2}a(q^2)\pi\frac{M_\rho^3}{\Gamma_\rho}\delta(s'-M_\rho^2) \,,
\label{eq:narrowwidth}
\end{align}
with $M_\rho$ ($\Gamma_\rho$) the mass (width) of the $\rho(770)$, and $a(q^2)$ as given in~\eqref{eq:BW} and~\eqref{eq:A}. 
Based on~\eqref{eq:DRunsub} together with
\beq
\Gamma_{\rho}=\frac{g_{\rho\pi\pi}^2}{6\pi}\frac{q_\pi^3(M_\rho^2)}{M_\rho^2}
\eeq
as well as the KSFR relation $2F_\pi^2g_{\rho\pi\pi}^2=M_\rho^2$~\cite{KSFR_1,KSFR_2},
this leads to the VMD-type approximation
\beq
F_{\pi^0\gamma^*\gamma^*}(q_1^2,q_2^2)=\frac{3M_\rho^2}{2}\frac{F_{\pi\gamma\gamma}}{F_{3\pi}}
\bigg\{\frac{a(q_2^2)}{M_\rho^2-q_1^2}+\frac{a(q_1^2)}{M_\rho^2-q_2^2}\bigg\}
\eeq
for the full doubly-virtual transition form factor. Ignoring the quark-mass renormalization of $F_{3\pi}$, this representation automatically satisfies the normalization condition
\beq
F_{\pi^0\gamma^*\gamma^*}(0,0)=F_{\pi\gamma\gamma}
\eeq
and predicts for the slope of the singly-virtual form factor
\beq
a_\pi=M_{\pi^0}^2\bigg(\frac{1}{2M_\rho^2}+\frac{3\beta}{2F_{3\pi}}\bigg).
\eeq
The VMD formula
\beq
\label{slope_VMD}
a_\pi^\text{VMD}=\frac{M_{\pi^0}^2}{M_V^2}\sim 30.7\times 10^{-3}
\eeq
for $M_V\sim 0.77\GeV$ is reproduced if we identify $M_\rho\to M_V$ and 
\beq
\beta\to\beta^\text{VMD}=\frac{F_{3\pi}}{3M_V^2}\sim 5.5\GeV^{-5},
\eeq
indeed close to the fit results shown in Table~\ref{tab:3pifit}. Analytically, this value of $\beta$ is reproduced by writing down an unsubtracted version of~\eqref{eq:BW}, employing a narrow-width approximation for the $\omega$, neglecting the $\phi$, and fixing the normalization to the chiral anomaly
\beq
a^\text{VMD}(q^2)=\frac{F_{3\pi}}{3}\frac{M_V^2}{M_V^2-q^2}=\frac{F_{3\pi}}{3}+\beta^\text{VMD}q^2+\mathcal{O}\big(q^4\big).
\eeq
The fact that our prediction for the slope~\eqref{slope} coincides exactly with the VMD value~\eqref{slope_VMD} is of course purely accidental:
already varying the VMD mass between the masses of $\rho$ and $\omega$ produces the interval $(29.7\ldots 30.3)\times 10^{-3}$ and therefore $a_\pi^\text{VMD}=30.0\times 10^{-3}$ if the physical masses are kept in the above derivation. In fact, the reason why the final number is much closer to the original VMD prediction can be attributed to the inclusion of the $\phi$, which in the narrow-width approximation 
leads to
\begin{align}
\label{VMD_improved}
a_\pi^\text{VMD}&\to \frac{M_{\pi^0}^2}{2}\bigg\{\frac{1}{M_\rho^2}+\frac{1}{1+c}\bigg(\frac{1}{M_\omega^2}+\frac{c}{M_\phi^2}\bigg)\Bigg\}\notag\\
&\sim 30.5\times 10^{-3},
\end{align}
with $c=c_\phi/c_\omega\times M_\omega^2/M_\phi^2\sim -0.08$
from Table~\ref{tab:3pifit}. 
We also note that our prediction for $b_\pi$ given in~\eqref{slope2} is {\it not} consistent with the VMD value
\beq
b_\pi^\text{VMD}=\frac{M_{\pi^0}^4}{M_V^4}\sim 0.94\times 10^{-3},
\eeq
even the more realistic analog of~\eqref{VMD_improved} produces  $b_\pi\sim 0.93\times 10^{-3}$.
All these considerations show that, while the general aspects can be understood in the VMD framework, reliable uncertainty estimates require a full calculation as presented in this paper.

\end{document}